\pdfoutput=1
\documentclass[11pt]{article}
\usepackage[final]{acl}
\usepackage{times}
\usepackage{latexsym}

\usepackage[T1]{fontenc}
\usepackage[utf8]{inputenc}

\usepackage[final,nopatch=footnote]{microtype}
\usepackage{inconsolata}
\usepackage{graphicx}
\usepackage[ruled,vlined,lined,commentsnumbered]{algorithm2e}
\usepackage{amsmath,amsfonts}
\usepackage{amssymb}
\usepackage{array}
\usepackage{booktabs}
\usepackage[skip=1pt,labelfont=bf]{caption}
\usepackage{calligra}
\usepackage{color, colortbl}
\usepackage{courier}
\usepackage{csvsimple}
\usepackage{enumitem}
\usepackage{fancybox}
\usepackage{listings}
\usepackage{longtable}
\usepackage{lscape}
\usepackage{makecell}
\usepackage{marvosym}
\usepackage{moreverb}
\usepackage{multicol}
\usepackage{multirow}
\usepackage{pifont}
\usepackage{rotating}
\usepackage{setspace}
\usepackage{caption}
\usepackage{subfigure}
\usepackage{parcolumns}
\usepackage[most]{tcolorbox}
\usepackage{threeparttable}
\usepackage{tikz}
\usepackage[normalem]{ulem}
\usepackage{url}
\usepackage{soul}
\usepackage{wasysym}
\usepackage{svg}
\usepackage{textcomp}
\usepackage{xcolor}
\usepackage{wrapfig}
\usepackage{pdflscape}
\usepackage{hyperref}
\usepackage{tikz}

\usepackage{amsmath}
\usepackage{array}
\usepackage{balance}
\colorlet{punct}{red!60!black}
\definecolor{background}{HTML}{EEEEEE}
\definecolor{delim}{RGB}{20,105,176}
\definecolor{yellow}{RGB}{254, 254, 0}
\definecolor{lightblue}{RGB}{0, 254, 254}
\definecolor{mygray}{gray}{0.98}
\colorlet{numb}{magenta!60!black}

\lstdefinestyle{sol}{
    language=Python,
    basicstyle=\ttfamily\small,
    keywordstyle=\color{blue!70},
    commentstyle=\color{green!50!black},
    stringstyle=\color{red!70},
    numbers=left,
    numberstyle=\tiny\color{gray},
    stepnumber=1,
    showspaces=false,
    showstringspaces=false,
    breaklines=true,
    frame=single,
    tabsize=4,
    morekeywords={function, address, external, returns, internal, assertEq},
    escapeinside=`` 
}

\lstdefinelanguage{json}{
    basicstyle=\normalfont\ttfamily,
    numbers=left,
    numberstyle=\scriptsize,
    stepnumber=1,
    numbersep=8pt,
    showstringspaces=false,
    breaklines=true,
    frame=lines,
    backgroundcolor=\color{background},
    literate=
     *{0}{{{\color{numb}0}}}{1}
      {1}{{{\color{numb}1}}}{1}
      {2}{{{\color{numb}2}}}{1}
      {3}{{{\color{numb}3}}}{1}
      {4}{{{\color{numb}4}}}{1}
      {5}{{{\color{numb}5}}}{1}
      {6}{{{\color{numb}6}}}{1}
      {7}{{{\color{numb}7}}}{1}
      {8}{{{\color{numb}8}}}{1}
      {9}{{{\color{numb}9}}}{1}
      {:}{{{\color{punct}{:}}}}{1}
      {,}{{{\color{punct}{,}}}}{1}
      {\{}{{{\color{delim}{\{}}}}{1}
      {\}}{{{\color{delim}{\}}}}}{1}
      {[}{{{\color{delim}{[}}}}{1}
      {]}{{{\color{delim}{]}}}}{1},
}

\newcommand{\mytitle}{SolEval\xspace}
\newcommand{\datasetname}{SolEval\xspace}

\titlebox{6cm}
\title{SolEval: Benchmarking Large Language Models for Repository-level Solidity Code Generation}

\author{Zhiyuan Peng$^{1,2}$\thanks{Equal contribution.}  \quad Xin Yin$^3$\footnotemark[1] \quad Rui Qian$^{4}$ \quad Peiqin Lin$^{5,6}$ \\ \bf Yongkang Liu$^{7}$ \quad Hao Zhang$^{8}$ \quad Chenhao Ying$^{1}$\thanks{Corresponding authors.} \quad Yuan Luo$^{1}$\footnotemark[2] \quad \\
$^{1}$ Shanghai Jiao Tong University, China \quad
$^{2}$ Shanghai SimMed Technology Co, Ltd \\
$^{3}$ Zhejiang University, China \quad
$^{4}$ Fudan University, China \\
$^{5}$ LMU Munich, Germany \quad
$^{6}$ Munich Center for Machine Learning, Germany \\
$^{7}$ Northeastern University, China \quad 
$^{8}$ Universiti Sains Malaysia, Malaysia
\\
\texttt{\{pzy2000,yingchenhao,yuanluo\}@sjtu.edu.cn} \\
\texttt{xyin@zju.edu.cn} \quad
\texttt{qianruii@126.com} \\
\texttt{linpq@cis.lmu.de} \quad 
\texttt{misonsky@163.com} \quad \texttt{zhanghao666@student.usm.my} \\
}

\begin{document}
\maketitle

\begin{abstract}
Large language models (LLMs) have transformed code generation.
However, most existing approaches focus on mainstream languages such as Python and Java, neglecting the Solidity language, the predominant programming language for Ethereum smart contracts.
Due to the lack of adequate benchmarks for Solidity, LLMs' ability to generate secure, cost-effective smart contracts remains unexplored.
To fill this gap, we construct \mytitle, the first repository-level benchmark designed for Solidity smart contract generation, to evaluate the performance of LLMs on Solidity.
\mytitle consists of 1,507 samples from 28 different repositories, covering 6 popular domains, providing LLMs with a comprehensive evaluation benchmark.
Unlike the existing Solidity benchmark, \mytitle not only includes complex function calls but also reflects the real-world complexity of the Ethereum ecosystem by incorporating Gas@k and Vul@k.
We evaluate 16 LLMs on \mytitle, and our results show that the best-performing LLM achieves only 26.29\% Pass@10, highlighting substantial room for improvement in Solidity code generation by LLMs.
Additionally, we conduct supervised fine-tuning (SFT) on Qwen-7B using SolEval, resulting in a significant performance improvement, with Pass@5 increasing from 16.67\% to 58.33\%, demonstrating the effectiveness of fine-tuning LLMs on our benchmark.
We release our data and code at \url{https://github.com/pzy2000/SolEval}.
\end{abstract}

\section{Introduction}
The rapid expansion of blockchain technology and Decentralized Finance (DeFi) has led to a significant surge in smart contract deployments. This growth brings about increased development pressures and elevated security demands, highlighting the critical need for efficient and reliable Solidity code generation tools~\cite{qian2023empirical,chaliasos2024smart}. As the cornerstone of Ethereum contracts, Solidity plays a fundamental role in enabling the decentralized applications that are driving the blockchain revolution~\cite{smaragdakis2025program}.

\begin{figure}[t]
\centering
\includegraphics[width=\columnwidth]{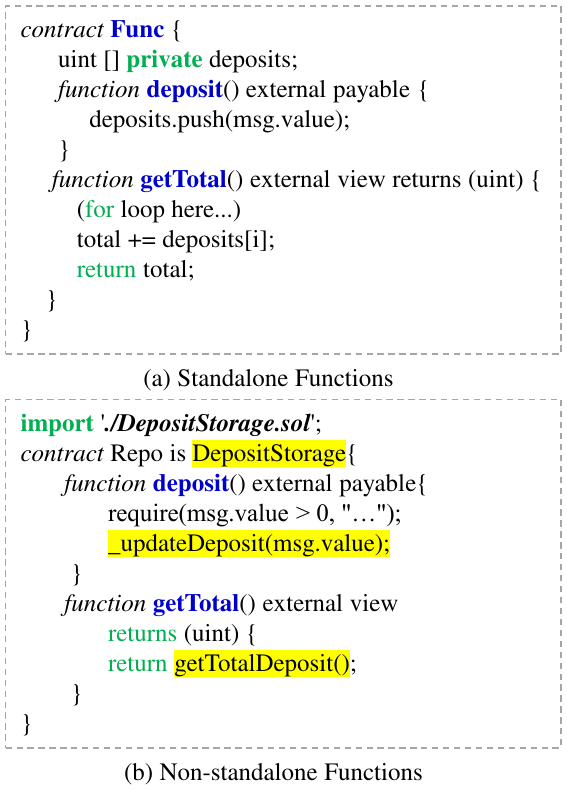}
\caption{Examples of standalone and non-standalone functions in Solidity with highlighted context dependencies. 
Repository-level code generation usually contains non-standalone function generation.}
\label{fig:example}
\end{figure}

\begin{table*}[htbp]
  \centering
  \caption{Comparison of existing benchmarks and \mytitle. Sample: number of class/function samples. SA Ratio: ratio of standalone functions. Dependency: number of dependencies (e.g., cross-file invocations). Avg. Token: average tokens in function requirements. Repo-Level: whether the benchmark is repository-level or not.}
  \small
    \begin{tabular}{lrrrrrrc}
    \toprule
    Benchmark & Sample & SA Ratio & Dependency & File & Avg. Token & Language & Repo-Level \\
    \midrule
    CoNaLa~\cite{yin2018learning} & 500   & 100\% & 0     & 0     & 13.1  & Python & \ding{55} \\
    HumanEval~\cite{chen2021evaluating} & 164   & 100\% & 0     & 0     & 58.8  & Python & \ding{55} \\
    MBPP~\cite{austin2021program}  & 974   & 100\% & 0     & 0     & 16.1  & Python & \ding{55} \\
    PandasEval~\cite{zan2022cert} & 101   & 100\% & 0     & 0     & 29.7  & Python & \ding{55} \\
    NumpyEval~\cite{zan2022cert} & 101   & 100\% & 0     & 0     & 30.5  & Python & \ding{55} \\
    AixBench~\cite{hao2022aixbench} & 175   & 100\% & 0     & 0     & 34.5  & Java  & \ding{55} \\
    ClassEval~\cite{du2023classeval} & 100   & 100\% & 0     & 0     & / & Python & \ding{55} \\
    Concode~\cite{iyer2018mapping} & 2,000 & 20\%  & 2,455 & 0     & 16.8  & Java  & \ding{51} \\
    CoderEval~\cite{yu2024codereval} & 230   & 36\%  & 256   & 71    & 41.5  & Python, Java & \ding{51} \\
    DevEval~\cite{li2024deveval} & 1,825 & 27\%  & 4,448 & 164   & 101.6 & Python & \ding{51} \\
    \midrule
    BenchSol~\cite{benchmark2024sol} & 15    & 100\% & 0     & 0     & 41.7 & Solidity & \ding{55} \\
    \cellcolor{lightgray}\textbf{SolEval} & \cellcolor{lightgray}1,507 & \cellcolor{lightgray}89\% & \cellcolor{lightgray}1,343 & \cellcolor{lightgray}129    & \cellcolor{lightgray}143.5 & \cellcolor{lightgray}Solidity & \cellcolor{lightgray}\ding{51} \\
    \bottomrule
    \end{tabular}
  \label{tab:Statistics_other_benchmarks}
\end{table*}

Recently, methods based on large language models (LLMs) have become the dominant approach to code generation~\cite{radford2018improving,brown2020language,yu2024codereval}. 
These methods can generate the corresponding functions according to descriptions in natural language.
To assess the code generation capabilities of models, researchers have proposed a series of benchmarks~\cite{du2023classeval,yu2024codereval,li2024deveval,benchmark2024sol}.
As shown in Table~\ref{tab:Statistics_other_benchmarks}, most of these benchmarks focus on mainstream programming languages such as Python and Java, with little attention paid to Solidity.
Different from the high flexibility of programming languages like Python, Solidity's operation is constrained by gas fee (costs of executing operations on a blockchain) and blockchain immutability, making Solidity code generation more challenging than general programming languages.
To evaluate the coding abilities of LLMs in Solidity, \citet{benchmark2024sol} proposes the first Solidity benchmark, BenchSol.
However, BenchSol is entirely generated by GPT-4, distinct from real-world scenarios. 
Moreover, this benchmark is severely limited in scale, featuring only 15 functions, and is restricted to standalone functions (i.e., Non-repository-level generation). Cross‑file calls, library imports, and storage layout, which are absent in Non-repository-level benchmarks, directly reflect the complexity of real-world smart contract development.

To fill the gap, we propose~\mytitle, the first benchmark that supports repository-level smart contract generation.
As shown in Figure~\ref{fig:example}, \mytitle contains non-standalone functions that invoke context dependencies from other files, which are absent in the existing Solidity benchmark.
\ding{182} \mytitle contains 1,507 samples from 28 real-world repositories, covering 6 popular domains (e.g., security, economics, and games).
\ding{183} \mytitle is manually annotated by 5 master's students with Solidity experience. 
\mytitle contains detailed requirements, repositories, codes, context information, and test cases.
\ding{184} To evaluate secure and cost-effective smart contract generation, we incorporate Gas@k and Vul@k attributes into \mytitle.

We evaluate 16 popular LLMs on \mytitle, including closed-source models (e.g., GPT-4o and GPT-4o-mini) and open-source models (e.g., CodeLlama and DeepSeek-R1). The results reveal a striking performance gap: these models achieve a Pass@10 ranging from 5.91\% to 26.29\%, indicating that their performance in Solidity code generation is far from optimal, with significant room for improvement. The generated smart contracts exhibit varying gas fees and vulnerability rates, highlighting the dilemma of balancing cost efficiency with security in contract generation.

We also have an interesting finding: DeepSeek-V3 ranks highest in Pass@10 but generates contracts with high gas fees, while DeepSeek-R1-Distill-Qwen-7B ranks lowest but generates the cheapest contracts. This contrast highlights a fundamental challenge in Solidity code generation: balancing functional correctness with gas efficiency. LLMs excelling in generating correct code may struggle with optimizing gas costs, while models focused on optimizing gas efficiency may sacrifice the quality or correctness of the generated code.

\begin{figure*}[htbp!]
    \centering
    \includegraphics[width=0.96\linewidth]{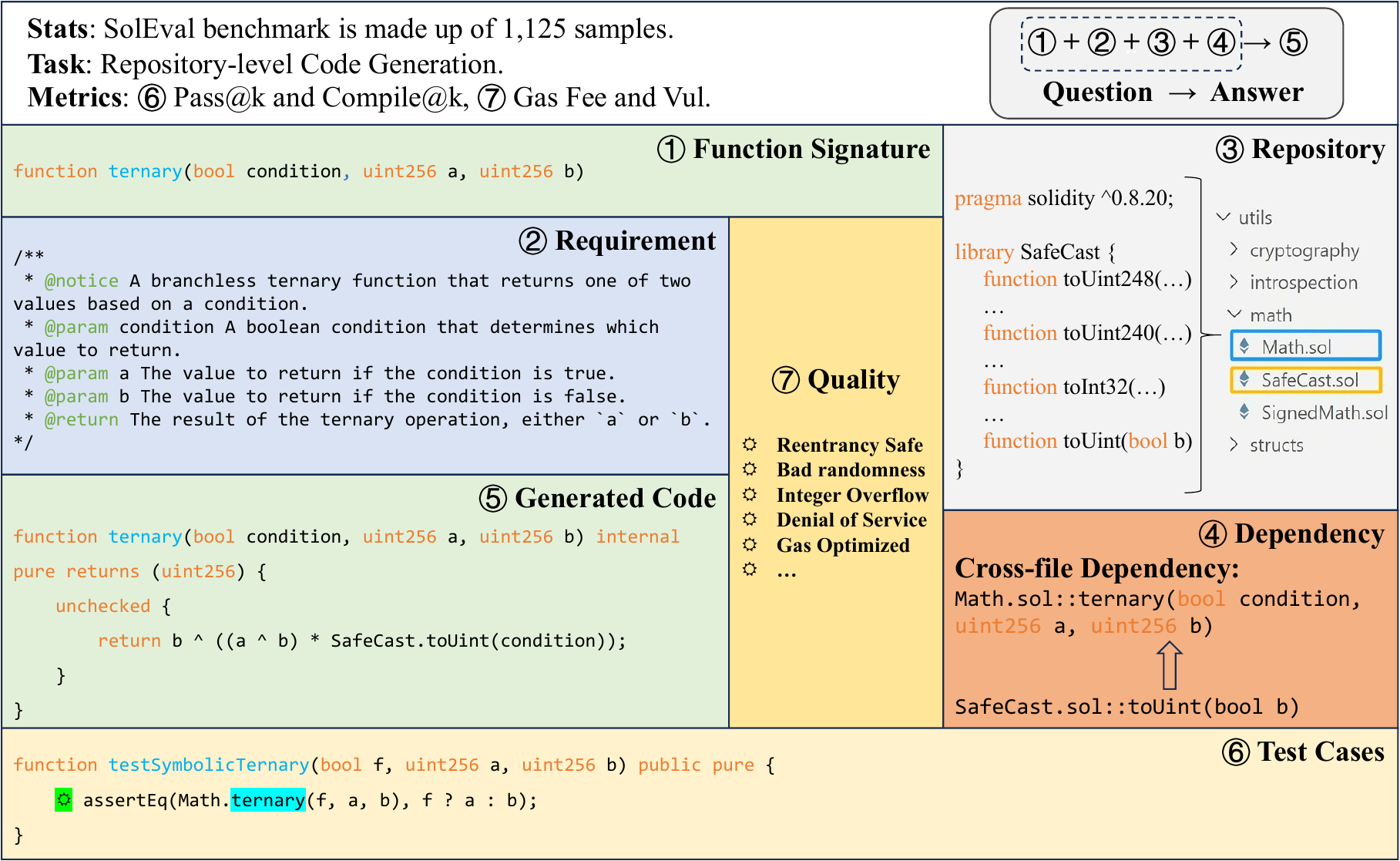}
    \caption{Overview of the \datasetname benchmark for Solidity code generation.}
    \label{fig:overview}
\end{figure*}

Additionally, we discover that the inclusion of Retrieval-Augmented Generation (RAG) and contextual information improves model performance, highlighting the importance of incorporating contextual awareness in Solidity code generation tasks. In particular, we conduct supervised fine-tuning (SFT) on Qwen-7B using SolEval, resulting in a significant performance improvement. Pass@5 increases from 16.67\% to 58.33\%, demonstrating that fine-tuning LLMs on our benchmark leads to a notable enhancement in the generation of high-quality Solidity code. This reinforces the effectiveness of our benchmark in improving LLM performance through task-specific training.

In summary, our contributions are as follows:

\begin{itemize}[leftmargin=*]
    \item We introduce the first repository-level benchmark for Solidity code generation, including a diverse set of 1,507 samples from 28 real-world repositories, covering 6 popular domains. 
    We also propose essential metrics (i.e., Gas@k and Vul@k) critical for smart contract development.
    \item We conduct an extensive evaluation of 16 state-of-the-art LLMs on \mytitle, revealing their performance gaps when generating smart contracts. We find that LLMs can generate better contracts when using RAG and context information.
    \item We conduct supervised fine-tuning (SFT) on Qwen-7B using SolEval, demonstrating a significant performance improvement, with Pass@5 increasing from 16.67\% to 58.33\%. This substantial enhancement highlights the effectiveness of fine-tuning LLMs on our benchmark for generating high-quality Solidity smart contracts.
\end{itemize}

\label{sec:introduction}

\section{Benchmark - \datasetname}

\subsection{Overview}

\datasetname contains 1,507 samples from 28 real-world code repositories, covering 6 popular domains (e.g., security, economics, and games).
The statistics for the projects are shown in Table~\ref{tab:dataset-statistics}. 
The functions that are filtered out can still serve as knowledge databases for RAG to select examples.

\begin{table}[htbp!]
\centering
\caption{The simplified statistics of the top-9 projects. 
Fi.: Filtered Functions with rules defined in Section~\ref{sec:project_select}.
}
\label{tab:dataset-statistics}
\resizebox{\linewidth}{!}
{
\begin{tabular}{lccc}
\toprule
\textbf{Top 9 Project} & \textbf{Function} & \textbf{Test Case} & \textbf{LOC} \\
\midrule
\href{https://github.com/Vectorized/solady}{Solady} & 4,570 &  1,389 & 9.68 \\
% Openzeppelin-contracts
\href{https://github.com/OpenZeppelin/openzeppelin-contracts}{Contracts}
& 2,453  & 217 & 7.39 \\
\href{https://github.com/OpenZeppelin/ethernaut}{Ethernaut} & 445 & 86 & 6.10 \\
% Openzeppelin-foundry-upgrades
\href{https://github.com/OpenZeppelin/openzeppelin-foundry-upgrades}{foundry-upgrades} & 5,317 & 70 & 4.70 \\
\href{https://github.com/0xethsign/Account2}{Account2} & 13 & 2 & 6.93 \\
% Openzeppelin-community-contracts
\href{https://github.com/OpenZeppelin/openzeppelin-community-contracts}{community-contracts} & 1,372 & 12 & 3.77 \\
% Openzeppelin-contracts-upgradeable
\href{https://github.com/OpenZeppelin/openzeppelin-contracts-upgradeable}{contracts-upgradeable}& 1,663 & 161 & 4.53 \\
% Uniswap-solidity-hooks-template
\href{https://github.com/OpenZeppelin/uniswap-solidity-hooks-template}{Uniswap-solidity} & 39 & 10 & 15.8 \\
\href{https://github.com/foundry-rs/forge-std}{Forge-std} & 1,951 & 270 & 8.66 \\
\midrule
Total & 17,823 (Fi.: 1,125) & 2,217 & 6.76 \\
\bottomrule
\end{tabular}
}
\end{table}

\datasetname benchmarks LLMs on repository-level smart contract generation, consisting of two phases: (1) LLM-based Solidity Code Generation (\S\ref{sec:llm_cg}) and (2) Post-Generation Evaluation (\S\ref{sec:eval_metrics}).

As illustrated in Fig.~\ref{fig:overview}, the first phase involves the evaluated LLM taking a function signature, requirements, and repository dependencies as input (\ding{182}\ding{183}\ding{184}\ding{185}). 
The LLM then generates a function (\ding{186}) that satisfies the specified requirements.
In the Post-Generation Evaluation phase, the generated function is integrated into the repository to get the generated smart contract, and its functional correctness (\ding{187}) and quality attributes (\ding{188}) are evaluated.

\subsection{LLM-based Solidity Code Generation}
\label{sec:llm_cg}

The evaluated LLM receives the following inputs: \ding{182} \textbf{Function Signature}: The function's signature. \ding{183} \textbf{Requirement}: A natural language description of the function, also referred to as `comment' in later sections. \ding{184} + \ding{185} \textbf{Repository Context}: Code contexts (e.g., interfaces, functions, variables) defined outside the target code and invoked in the reference code. 
The LLM is then prompted (see \S\ref{sec:prompt} for details) to generate a desired function, which is subsequently injected into the repository to get the smart contract for real-world code evaluation.

\subsection{Post-Generation Evaluation}
\label{sec:eval_metrics}

Following \citet{britikov2024soltg}, we utilize an executor that verifies functional correctness, accommodating differences across Solidity compilers and handling unit test distribution, to execute the test cases. We evaluate functional correctness (\ding{187}) using Pass@k and Compile@k, and assess quality attributes (\ding{188}) with Gas@k and Vul@k. See \S\ref{sec:passk}, \S\ref{sec:compk},  \S\ref{sec:vulk} and \S\ref{sec:gask} for detailed definitions.

\section{Benchmark Construction}

\begin{figure}[htbp!]
    \centering
    \includegraphics[width=\linewidth]{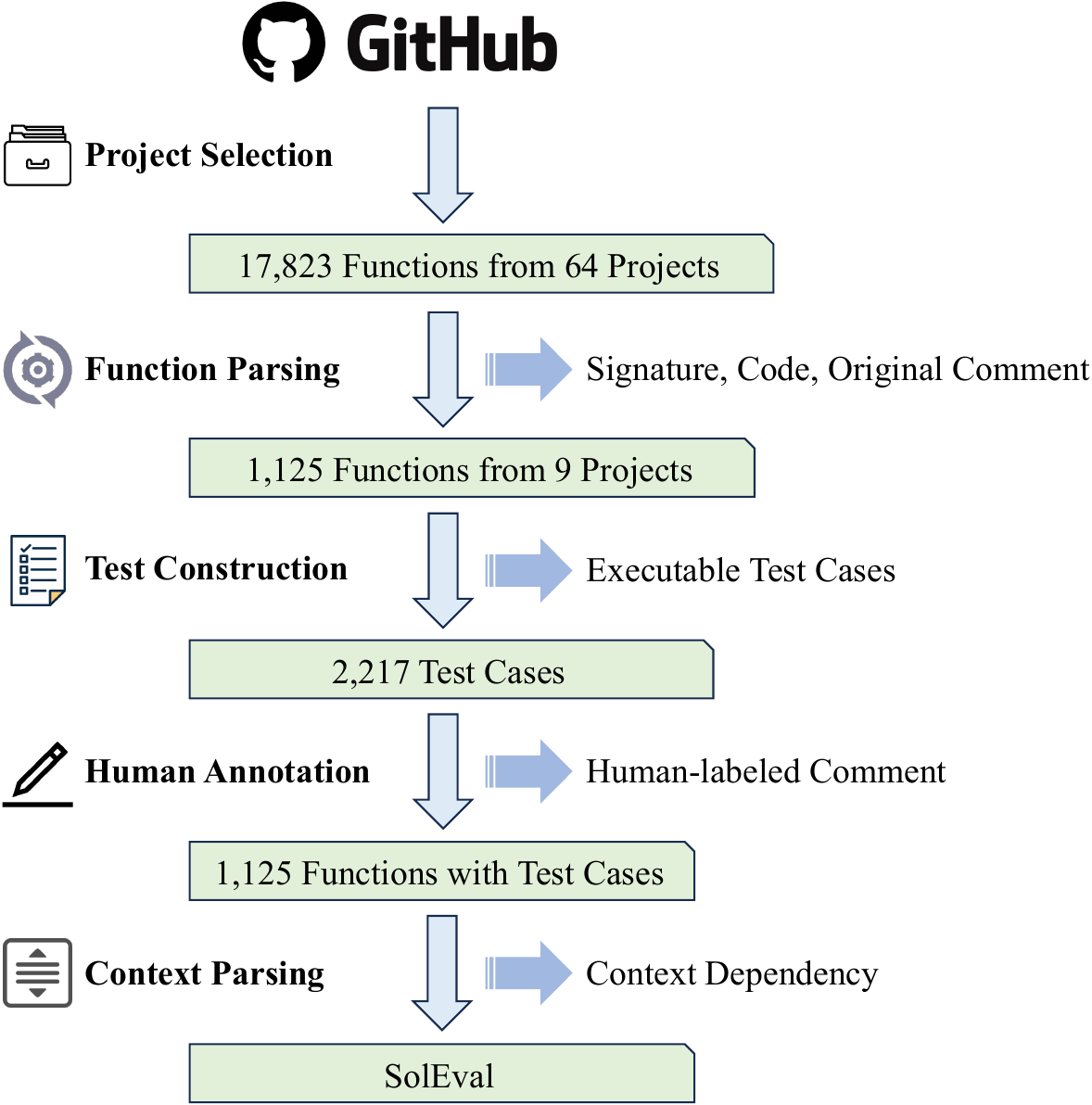}
    \caption{The process of constructing \datasetname.}
    \label{fig:collection}
\end{figure}

As shown in Fig.~\ref{fig:collection}, the construction of \datasetname involves five key phases.

\subsection{Project Selection}
\label{sec:project_select}
% \textbf{Phase \ding{182}: Project Selection.}
To ensure \datasetname's practicality and diversity, we follow best practices~\cite{chen2021evaluating,yu2024codereval,liu2024your} and select functions from different open-source projects through four steps. 
First, we manually select six popular GitHub organizations, such as OpenZeppelin, that host Solidity projects.
We crawl all their public repositories, sort them by star count in descending order, and filter out low-star (i.e., with fewer than 40 stars) projects lacking test cases or containing fewer than 10\% files written in the Solidity language. By manually selecting popular GitHub projects, we ensure that \datasetname assesses a model’s ability to generate smart contracts that are more likely to be used within the blockchain community.

We then select functions that may be used in real scenarios based on three criteria: 
(1) We exclude trivial functions with fewer than five lines of code (LOC), following previous studies \citep{tse24gassmell}; 
(2) We exclude functions that are rarely deployed in real-world scenarios, as assessed by five master's students. 
Given that developers may have varying preferences regarding frequently used functions, the inclusion of a diverse set of preferences helps mitigate potential bias; 
and (3) We exclude test functions or deprecated functions.

\subsection{Function Parsing}
We extract all functions from the selected projects.
Since native Tree-sitter~\cite{treesitter} support for Solidity is inadequate for use, we design a Solidity version of Tree-sitter to accurately parse Solidity contracts and extract relevant information (e.g., function identifiers, bodies, and requirements).
From the extracted functions, we filter out tests, interfaces, and functions with LOC smaller than five, and retain those functions invoked by test functions, successfully compiled, and passed the original test cases.
This process results in 1,125 function samples from different Solidity projects.

\subsection{Test Construction}
To enhance the reliability of the evaluation, we take meticulous steps to ensure the correctness and completeness of the tests.
First, we analyze and collect the unit tests included in the project. 
For tests that did not provide sufficient line or branch coverage, we manually wrote additional test cases to ensure full line and branch coverage for the functions.

To ensure the correctness of the assessment of the generated functions, we employ advanced testing techniques (i.e., Fuzz, Invariant, and Differential Testing) using Forge~\cite{Foundry_Invariant_Test}.
To reproduce our gas fee result, it is suggested that the fuzzing seed is set to 666.

To establish a mapping between the focal functions and their corresponding test cases, we follow \citet{nie2023learning} and select the last function call before the first assertion from the test case.
Therefore, we identify the test cases for each focal function. 
This method minimizes the number of test cases per function. 
Evaluating the correctness of a function typically requires executing all test cases, which can be time-consuming. 
Consequently, in our experiment, we execute only the test cases that directly or indirectly call the target function, thereby reducing the testing time while maintaining comprehensive test coverage.

\subsection{Human Annotation}
Prompts play a crucial role in the performance of LLMs~\cite{jang2023can,sarkar2022like,shrivastava2023repository,zhou2022learning,zhou2022large}. 
In code generation tasks, the quality of the generated code is significantly influenced by the input requirements. 
Function-level comments serve multiple purposes, including explaining internal logic, describing behaviour and external usage, and stating effects and precautions~\cite{yu2024codereval}.

We recruit five master's students with at least three years of Solidity experience to provide double-checked, manually annotated function descriptions. 
There are two reasons for incorporating manually annotated comments into \datasetname: (1) to reduce the LLMs' memorization effects, as original comments are highly likely to have been encountered during the pre-training phase, and (2) to provide high-quality comments for the functions in \datasetname. 
To ensure the quality and consistency of the annotated function descriptions, we perform an inter-annotator agreement analysis using Fleiss' Kappa~\cite{fleiss1971measuring}. 
The classification of annotations into four categories (intact, partially intact, unclear, and unlabeled) was performed manually by annotators through the following steps: (1) Each function was independently annotated by two annotators; (2) Disagreements were resolved through a discussion moderated by a third expert annotator; (3) Inter-annotator reliability was evaluated using Fleiss' Kappa to ensure high-quality and consistent annotations. 
By calculating the observed agreement (\(P_o\)) and the expected agreement (\(P_e\)) under the assumption of independent classifications, Fleiss' Kappa serves as a reliable indicator of annotator alignment, ranging from complete agreement (\(\kappa = 1\)) to random agreement (\(\kappa = 0\)). We consider \(\kappa = 0.8\) an excellent level of agreement, indicating that our annotators' decisions are highly consistent.

\subsection{Context Parsing}
One of the key differences between \mytitle and existing benchmark~\cite{benchmark2024sol} is our consideration of contextual dependencies. 
In repository-level code generation, a token undefined error often occurs when the necessary context is missing, leading to compilation errors~\cite{A3CodGen}. 
Therefore, providing relevant context (e.g., function signatures) is essential to help \mytitle validate the model's understanding of the requirement.

To maintain efficiency and avoid unnecessary costs or performance degradation, it is crucial to ensure that the contextual information is concise~\cite{A3CodGen}. 
Following~\cite{yu2024codereval}, we define the context code (e.g., functions, variables, and interfaces) required by a function to execute as its contextual dependencies.
We identify the contextual dependencies of a function through a two-step program analysis of the entire project.
First, given a function to analyze, we retrieve the corresponding source file from the database and then parse it to obtain a list of type, function, variable, and constant definitions. 
Next, we use static program analysis to identify all external invocations defined outside the current function, retrieving the signatures of these invocations. 
We then store these invocation signatures along with other relevant information about the function sample.
\label{sec:benchmark}

\section{Experimental Setup}
We conduct the first study to evaluate existing LLMs on repository-level Solidity code generation by answering the following research questions:

\begin{itemize}[leftmargin=*]
\item \textbf{RQ-1 Overall Correctness.} {\em How do LLMs perform on Solidity code generation?}

\item \textbf{RQ-2 Sensitivity Analysis.} {\em How do different configurations affect the effectiveness of LLMs?}

\end{itemize}

\subsection{Studied LLMs}

We select 16 state-of-the-art LLMs widely used in recent code generation studies~\cite{khan2023xcodeeval,yan2023codescope,A3CodGen,yu2024codereval,li2024deveval}. 
In particular, we focus on recent models released since 2022, and we exclude the small models (with fewer than 2B parameters) due to their limited efficacy. 
Table~\ref{tab:studied_llm} presents the state-of-the-art LLMs studied in our experiments with their sizes and types. 
Our study includes a wide scope of LLMs that are diverse in multiple dimensions, such as (i) being both closed-source and open-source, (ii) covering a range of model sizes from 6.7B to 671B, (iii) being trained for general or code-specific purposes. 
For detailed descriptions of each model, refer to \S\ref{sec:base_llms}.

\begin{table}[htbp]
    \centering
    \caption{Overview of the studied LLMs}
    \resizebox{\linewidth}{!}
    {
        \begin{tabular}{llc}
        \toprule
        \textbf{Type} & \textbf{Name} & \textbf{Size} \\
        \midrule
        \multirow{6}[0]{*}{\textbf{General LLM}}& DeepSeek-V3 & 671B (API) \\
        & DeepSeek-R1-Distill-Qwen & 7B / 32B  \\
        & DeepSeek-R1-Distill-Llama & 8B \\
        & GPT-4o & - \\
        & GPT-4o-mini & - \\
        & QwQ & 32B \\
        \midrule
        \multirow{6}[0]{*}{\textbf{Code LLM}} & CodeLlama & 7B / 34B \\
        & DeepSeek-Coder & 6.7B / 33B \\
        & DeepSeek-Coder-V2-Lite & 16B \\
        & Magicoder-S-DS & 6.7B \\
        & OpenCodeInterpreter-DS & 6.7B \\
        & Qwen2.5-Coder & 7B / 32B \\
        \bottomrule
        \end{tabular}
    }
  \label{tab:studied_llm}
\end{table}

\subsection{Evaluation Methodology and Metrics}
We adopt the Pass@K and propose the Compile@K. 
The detailed explanations of the metrics are in \S\ref{sec:eval_metrics}.
We set the total number (denoted as n) of samples generated by an LLM to 10, and then calculate Pass@K for the LLM with K’s value of 1, 5, and 10, respectively, which is also the case for Compile@K.
When k = 1, we use the greedy search and generate a single program per requirement. 
When k > 1, we use the nucleus sampling with a temperature of 1 and sample k programs per requirement.
We set the top-p to 0.95 and the max generation length to 512.
We also propose Vul@k and Gas@k metrics.
The detail of these metrics is illustrated in \S\ref{sec:eval_metrics} and the Appendix~\S\ref{sec:base_llms}.
We follow \citet{parvez2021retrieval,chen2024code,yin2024thinkrepair} and use RAG to select the best examples and collect a database from our projects for RAG based on the functions excluded from \mytitle. 
For detailed descriptions of RAG, refer to \S\ref{RAG attributes}. Note that all experimental results are averaged over five independent runs. 

\subsection{Setup for Supervised Fine-Tuning}
To prepare the data for supervised fine-tuning of Qwen-7B, we first evaluated 16 LLMs on SolEval. We removed the generated patches that failed the unit tests and merged the remaining valid patches with the original SolEval dataset. This process resulted in a set of NL-Code pairs, where each pair consists of a natural language description and a corresponding code patch. We then split these NL-Code pairs into a training and validation set with a 9:1 ratio. For the SFT process, we used the training set to fine-tune Qwen-7B with a maximum input length of 2048 tokens. The model was trained for 3 epochs, with validation performed at the end of each epoch. All other hyperparameters were kept at the default values provided by the TRL library.

We chose Qwen-7B for SFT due to its strong performance in initial evaluations, making it a promising candidate for further fine-tuning. To prevent data leakage, we ensured that there were no overlapping functions between the training and test sets (i.e., no identical function bodies). We randomly selected 30 repositories from GitHub, excluded 9 repositories that contained potential data leakage, and used the remaining repositories as the test set.

\label{sec:experiment}

\begin{table*}[htbp]
    \centering
    \caption{Performance of LLMs on \datasetname, evaluated using Pass@k, Compile@k, Vul@k, and Gas@k.
    The table presents results under the one-shot setting with RAG and Context. 
    Bold values indicate the highest performance in each respective column. Based on the mathematical definition of Gas@k, Gas@k is always smaller than Pass@k.
}
    \resizebox{\linewidth}{!}
    {
        \begin{tabular}{lc|ccc|ccc|c|c}
        \toprule
        LLMs & Size & Pass@1 & Pass@5 & Pass@10 & Compile@1 & Compile@5 & Compile@10 & Vul@1$\downarrow$ & Gas@1$\uparrow$ \\
        \midrule
        \multicolumn{10}{c}{\cellcolor{lightgray}6.7B to 16B} \\
        \midrule
        DeepSeek-R1-Distill-Qwen & 7B & 2.08\% & 4.50\% & 5.91\% & 6.37\% & 18.27\% & 26.29\% & \textbf{10.59\%} & 0.99\% \\
        DeepSeek-R1-Distill-Llama & 8B & 3.67\% & 6.95\% & 8.45\% & 8.78\% & 21.68\% & 29.04\% & 20.07\% & 1.67\% \\
        DeepSeek-Coder-Lite & 16B & \textbf{10.10\%} & 14.94\% & 16.79\% & \textbf{39.44\%} & \textbf{54.21\%} & \textbf{57.55\%} & 26.91\% & \textbf{4.31\%} \\
        DeepSeek-Coder & 6.7B & 8.39\% & 14.25\% & 16.68\% & 32.45\% & 50.74\% & 54.59\% & 23.17\% & 3.65\% \\
        CodeLlama & 7B & 5.15\% & 11.38\% & 14.26\% & 19.88\% & 43.05\% & 49.95\% & 25.00\% & 2.03\% \\
        Magicoder-S-DS & 6.7B & 7.26\% & 13.80\% & 16.68\% & 26.81\% & 48.77\% & 53.64\% & 24.33\% & 3.16\% \\
        OpenCodeInterpreter-DS & 6.7B & 7.05\% & 12.96\% & 15.66\% & 27.05\% & 48.71\% & 53.76\% & 27.08\% & 2.94\% \\
        Qwen2.5-Coder & 7B & 9.13\% & \textbf{15.28\%} & \textbf{17.44\%} & 33.31\% & 50.34\% & 54.44\% & 29.26\% & 4.11\% \\
        GPT-4o-mini & - & 7.18\% & 12.37\% & 14.69\% & 38.04\% & 53.18\% & 56.66\% & 34.01\% & 2.42\% \\
        \midrule
        \multicolumn{10}{c}{\cellcolor{lightgray}32B to 671B} \\
        \midrule
        DeepSeek-V3 & 671B & \textbf{21.72\%} & \textbf{24.99\%} & \textbf{26.29\%} & \textbf{53.35\%} & 57.57\% & 58.61\% & 26.61\% & \textbf{7.13\%} \\
        DeepSeek-R1-Distill-Qwen & 32B & 10.19\% & 17.06\% & 19.77\% & 31.99\% & 55.31\% & 61.31\% & 23.84\% & 3.89\% \\
        QwQ & 32B & 9.10\% & 16.74\% & 20.26\% & 48.33\% & 72.47\% & 76.65\% & 22.18\% & 3.68\% \\
        DeepSeek-Coder & 33B & 8.32\% & 15.57\% & 18.92\% & 29.35\% & 50.08\% & 55.39\% & 23.08\% & 3.48\% \\
        CodeLlama & 34B & 6.80\% & 13.52\% & 16.47\% & 24.59\% & 48.68\% & 54.80\% & 25.47\% & 2.75\% \\
        Qwen2.5-Coder & 32B & 13.46\% & 19.28\% & 21.44\% & 44.03\% & 55.53\% & 57.87\% & 24.52\% & 5.36\% \\
        GPT-4o & - & 12.96\% & 20.79\% & 23.70\% & 47.04\% & \textbf{58.45\%} & \textbf{60.74\%} & \textbf{21.50\%} & 4.51\% \\
        \bottomrule
        \end{tabular}%
    }
    \label{tab:rq1}%
\end{table*}

\section{Results}

\subsection{RQ-1 How do LLMs perform on Solidity code generation?}
\label{sec:rq1}

\noindent 
\textbf{Evaluation of Pass@k and Compile@k for generated code.}
Table~\ref{tab:rq1} presents the overall performance of state-of-the-art LLMs on \datasetname. 
Among the 6.7B-to-16B models, DeepSeek-Coder-Lite achieves the highest Pass@1 and Compile@1, surpassing other models. 
Notably, DeepSeek-R1-Distill-Qwen-7B, which claims comparable performance to ChatGPT-o1-mini on benchmarks such as LiveCodeBench and CodeForces~\cite{deepseekr1}, underperforms compared to CodeLlama-7B. 
This discrepancy is likely due to DeepSeek-R1-Distill's lack of knowledge of Solidity, highlighting the importance of a specialized benchmark like \mytitle. 
Among the 32B-to-34B models, Qwen2.5-Coder outperforms others in both Pass@k and Compile@k.
Overall, DeepSeek-V3 performs best with a 26.29\% Pass@10.
It is noteworthy that the distilled version of DeepSeek-R1-Qwen-32B retains significantly more of the original model's Solidity code generation capabilities during distillation compared to its 7B counterpart.

\noindent 
\textbf{Evaluation of Gas (Fee/Gas@k) and Vulnerability Rate (Vul@k) for generated code.}
As shown in Table~\ref{tab:rq1}, there is a significant variation in gas fee and vulnerability rate across various LLMs. 
DeepSeek-V3 ranks first in Pass@k but generates the most gas-inefficient contracts among the 32B-to-671B models (The higher the fee, the less efficient the codes are).
Additionally, GPT-4o-mini, while being outperformed by GPT-4o in Pass@k and vulnerability rate, excels in generating contracts with lower gas fees.

\subsection{RQ-2 How do different configurations affect the effectiveness of LLMs?}
\label{sec:rq2}

\begin{figure}[htbp]
    \centering
    \includegraphics[width=\linewidth]{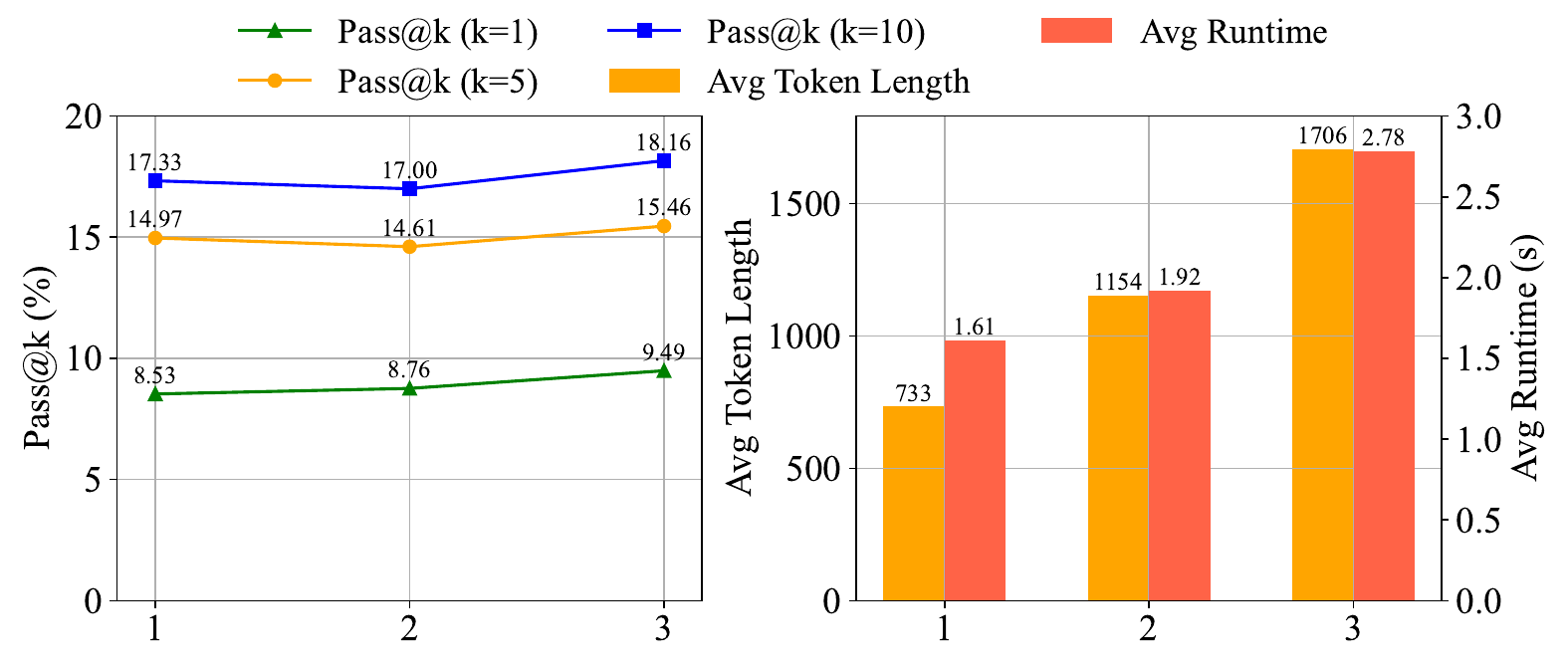}
    \caption{Performance of Qwen2.5-Coder-7B. The x-axis represents the number of shots.}
    \label{fig:ablation}
\end{figure}

\noindent 
\textbf{Impact of different example numbers.}
As previous studies~\cite{brown2020language,A3CodGen} have shown, the number of examples provided has a significant impact on LLMs' performance. 
To explore this, we adjust the number of examples while keeping other parameters and hyperparameters constant to ensure a fair comparison.
We do not conduct experiments in a zero-shot setting, as LLMs may generate unnormalized outputs without a prompt template, which would hinder automated extraction. 
From Fig.~\ref{fig:ablation}, we observe that as the number of examples increases, both the average token length and time cost rise sharply, while the improvement in Pass@k remains modest.
Based on these findings, we perform our ablation studies (Table~\ref{tab:rq1} and \ref{tab:ablation}) using a one-shot setting in \mytitle.

\noindent 
\textbf{Impact of different selection strategies.}
RAG retrieves relevant codes from a retrieval database and supplements this information for code generation~\cite{parvez2021retrieval}. 
To ensure a fair comparison, we set the number of examples to one and evaluated the results of RAG versus random selection on the same LLM (i.e., DeepSeek-V3). From Table~\ref{tab:ablation}, Pass@1 and Compile@1 are higher when RAG is enabled, indicating that it improves the effectiveness of code generation.

% \vspace{-0.2cm}

\noindent 
\textbf{Impact of Context Information.}
Since that relevant context typically enhances performance in other programming languages, we conduct an ablation study to examine the influence of context on the quality of LLM-generated contracts. Table~\ref{tab:ablation} shows that providing context information improves both Pass@1 and Compile@1. 
However, there is no clear correlation between gas fees, vulnerability rate, and the presence of context information. (See \S\ref{sec:gasfee} for introduction to gas fee)

\begin{table}[htbp]
    \centering
    \caption{Ablation study on the effect of RAG and Context on DeepSeek-V3's (one-shot) performance.}
    \resizebox{\linewidth}{!}
    {
        \begin{tabular}{cc|cccc}
        \toprule
        RAG & Context & Pass@1 & Compile@1 & Fee & Vul \\
        \midrule
        \ding{51} & \ding{51} & \textbf{21.72\%}& \textbf{53.35\%}&  \textbf{-7525}& 26.61\% \\ 
        \ding{55} & \ding{51} & 20.24\% & 51.08\% & 3828& \textbf{23.68\%}\\ 
        \ding{51} & \ding{55} & 21.28\% & 52.54\% & -708& 26.13\%\\
        \ding{55} & \ding{55} & 20.17\% & 50.32\% & 768& 26.83\%\\   
        \bottomrule
        \end{tabular}
    }
    \label{tab:ablation}
\end{table}

\subsection{Empirical Lessons}
\label{sec:Findings}

\paragraph{Supervised Fine-Tuning improves the Quality of the generated Solidity Codes.}
As shown in Table~\ref{tab:sft_performance}, SFT yields large gains across all metrics. Pass@5, Compile@5, and Gas@1 all improve substantially, while Vul@1 is reduced by over 19 percentage points. This confirms that supervised fine-tuning with SolEval boosts both correctness and robustness for Solidity code generation.

\begin{table}[htbp]
    \centering
    \caption{Performance of Qwen-7B before and after Supervised Fine-Tuning (SFT).}
    \resizebox{\linewidth}{!}
    {
        \begin{tabular}{c|ccccc}
        \toprule
        Strategies & Pass@5 & Compile@5 & Gas@1 & Vul@1 \\
        \midrule
        Before SFT & 16.67\% & 66.67\% & 0.00\% & 26.61\% \\
        \midrule
        After SFT & \textbf{58.83\%} & \textbf{100.00\%} & \textbf{19.84\%} & \textbf{7.35\%} \\
        \bottomrule
        \end{tabular}
    }
    \label{tab:sft_performance}
\end{table}

\paragraph{RAG and Context Information improve LLMs' performance in Solidity smart contract generation.}
As shown in Table~\ref{tab:ablation}, both Pass@1 and Compile@1 are higher when using RAG and context information.
This suggests that LLMs benefit from RAG and relevant contextual dependencies in generating more accurate and functional contracts. 
However, no significant correlation was observed between gas fee or vulnerability rate and the presence of context or RAG, indicating that while context and RAG enhance correctness, they do not necessarily influence efficiency or security.

\paragraph{While LLMs can generate pretty nice contracts with challenging requirements, they can fail in some really easy cases.} 
Fig.~\ref{fig:LLMs can generate pretty nice contracts} illustrates an example of GPT-4o solving a difficult requirement. 
On the other hand, Fig.~\ref{fig:LLMs can generate really dumb contracts} is an instance of DeepSeek-R1-Distill-Qwen-7B failing an easy problem. 
The detailed prompts and generated solutions are also provided in Fig.~\ref{fig:LLMs can generate pretty nice contracts} and Fig.~\ref{fig:LLMs can generate really dumb contracts}.

\paragraph{Larger language models improve the gas fee of the generated code.} 
Based on the data in Table~\ref{tab:rq1}, we observe that LLMs tend to generate more gas-efficient code. DeepSeek-V3 (671B) outperforms all other models in both Pass@k and gas efficiency, achieving the highest Pass@10 (26.29\%) and the best Gas@1 (7.13\%). Furthermore, the distilled version of DeepSeek-R1-Qwen (32B) maintains strong performance in Pass@k (19.77\% for Pass@10), while also demonstrating a notable improvement in gas efficiency compared to smaller models, with a Gas@1 score of 3.89\%. This suggests that larger models benefit from a stronger capacity to balance both functional correctness and gas efficiency in Solidity code generation.
\label{sec:results}

\section{Related Work}

\subsection{Large Language Model}

The advancement of pre-training technology has significantly advanced code generation in both academia and industry ~\cite{li2022competition,shen2022incorporating,nijkamp2022codegen,fried2023incoder}. 
This has led to the emergence of numerous Large Language Models (LLMs) that have made substantial strides in code generation, including ChatGPT~\cite{openai2022chatgpt}, Magicoder~\cite{wei2023magicoder}, CodeLlama~\cite{roziere2023code}, Qwen~\cite{bai2023qwen}, DeepSeek-Coder~\cite{deepseekcoder}, and OpenCodeInterpreter~\cite{zheng2024opencodeinterpreter}.

To optimize LLMs for various code generation scenarios, some previous studies focus on enhancing prompt engineering by introducing specific patterns, such as Structured Chain-of-Thought~\cite{yin2024thinkrepair,li2025structured}, Self-planning~\cite{jiang2024self}, Self-debug~\cite{chen2023teaching,xia2023keep}, and Self-collaboration~\cite{dong2024self}. 
However, these efforts primarily address mainstream programming languages (e.g., Java, Python, and C++)~\cite{yin2024rectifier,yin2024you,xia2023keep}.

\subsection{Code Generation Benchmark}
Existing benchmarks predominantly focus on mainstream programming languages (e.g., Python, Java), giving insufficient attention to Solidity language.

For mainstream languages, HumanEval is a widely recognized benchmark for evaluating code generation models on the functional correctness of code generated from docstrings~\cite{chen2021evaluating}. 
It consists of 164 hand-crafted programming problems, each with a corresponding docstring, solution in Python, function signature, body, and multiple unit tests. 
Following HumanEval, AiXBench~\cite{hao2022aixbench} was introduced to benchmark code generation models for Java. 
AiXBench contains 175 problems for automated evaluation and 161 problems for manual evaluation.
The authors propose a new metric to automatically assess the correctness of generated code and a set of criteria for manually evaluating the overall quality of the generated code. 
MultiPL-E~\cite{cassano2023multipl} is the first multi-language parallel benchmark for text-to-code generation. It extends HumanEval and MBPP~\cite{austin2021program} to support 18 programming languages.

While all the aforementioned benchmarks focus on standalone functions, DS-1000~\cite{lai2023ds} introduces non-standalone functions. 
It includes 1000 problems, covering seven widely used Python data science libraries, including NumPy, Pandas, TensorFlow, PyTorch, Scipy, Scikit-learn, and Matplotlib.
To mitigate data leakage, the authors manually modify functions and emphasize the use of real development data in DS-1000.

Concode~\cite{iyer2018mapping} is a large dataset containing over 100,000 problems from Java classes in open-source projects.
The authors collect Java functions with at least one contextual dependency from approximately 33,000 GitHub repositories. 
These functions are paired with natural language annotations (e.g., Javadoc-style method descriptions) and code.
The dataset is split at the repository level rather than the function level, and while it includes contextual dependencies, it uses BLEU as the sole evaluation metric and does not evaluate the correctness of the generated functions. 
Additionally, none of the above benchmarks supports Solidity.

For Solidity language, BenchSol~\cite{benchmark2024sol} is the only available benchmark for Solidity smart contract generation. 
It contains 15 use cases of varying difficulty levels and utilizes Slither and Hardhat.
However, BenchSol is hand-crafted, poorly aligned with real-world code repositories, and extremely limited in scale, only supporting the evaluation of standalone functions (i.e., Non-repository-level generation) for LLMs.
\label{sec:related_work}

\section{Conclusion and Future Work}

This paper presents a new benchmark named \datasetname to evaluate LLMs' effectiveness in Solidity smart contract generation scenarios.
Compared with BenchSol~\cite{benchmark2024sol}, \mytitle supports repository-level smart contract generation and excels in scale (75 times in number of functions) and real-world code alignment.
Meanwhile, our benchmark takes vulnerability rate and gas fee into consideration, both of which are crucial for secure and cost-effective smart contract development.
The experimental results show that SolEval can reveal the weaknesses of 16 state-of-the-art LLMs, highlighting the limitations of these LLMs in generating non-standalone Solidity functions.

In the future, there are two main directions for extending \mytitle. 
Firstly, we are looking for more high-quality code repositories from GitHub and enlarging \mytitle with more projects. 
Secondly, we plan to leverage SFT and DPO to fine-tune LLMs to generate safer and cheaper code.
\label{sec:conclusion}

\section*{Limitations}
We believe that \mytitle has four limitations:

\begin{itemize}[leftmargin=*]
    \item
    \mytitle is currently a monolingual benchmark, focusing solely on Solidity code generation.
    This approach overlooks the necessity for LLMs to comprehend requirements in various natural languages and to generate code in multiple programming languages, including Vyper and Rust.
    Recognizing this limitation, we plan to develop a multilingual version of \mytitle in future work to better assess LLMs' capabilities across diverse linguistic and programming contexts.
    \item
    Due to funding constraints, we were unable to evaluate \mytitle on GPT-5 and its competitors (e.g., Claude Opus 4.1) in our study. 
    This limitation may affect the generalizability of our findings, as these models have demonstrated advanced capabilities in various benchmarks.
    \item
    The gas fee and vulnerability rate metrics are limited to evaluating the gas efficiency and potential vulnerabilities of smart contracts without providing mechanisms for their optimization or remediation.
    In future work, we plan to extend our research to include methods for gas optimization and vulnerability detection (e.g., DPO for secure and gas-efficient solidity generation).
\end{itemize}

\section*{Ethics Consideration}
\datasetname is collected from real-world smart contract repositories. 
All samples in \datasetname are manually reviewed by five master's students, under the supervision of two PhD researchers in the field of code generation. 
We ensure that none of the samples contain private information or offensive content.

\section*{Acknowledgements}

This work was supported in part by National Key Research and Development Program of China under Grant 2024YFB2705300, in part by the Shanghai Science and Technology Innovation Action Plan under Grant 23511100400, in part by the National Natural Science Foundation of China under Grant 62402313, in part by the Open Research Fund of The State Key Laboratory of Blockchain and Data Security, Zhejiang University.
\label{sec:limitations}

\bibliography{main}

\clearpage
\appendix

\section{Glossary}
\begin{tcolorbox}[colback=mygray, colframe=black!20, arc=2mm, boxrule=0.3pt, left=1mm, right=1mm, top=1mm, bottom=1mm]
  \textbf{Blockchain:} A distributed ledger that records transactions across multiple computers in a way that ensures data integrity and security. \\[0.7em]
  \textbf{Smart contract:} A self-executing program stored on a blockchain (e.g., Ethereum) that automatically runs when predetermined conditions are met. \\[0.7em]
  \textbf{Solidity:} The primary programming language for writing Ethereum smart contracts. It is a statically-typed language with a syntax similar to JavaScript. \\[0.7em]
  \textbf{Repository-level code generation:} The task of generating code in the context of a software repository (project) rather than a single isolated function or file. \\[0.7em]
  \textbf{Gas fee:} The cost required to execute a transaction or operation on Ethereum, measured in units of ``gas''. 
\end{tcolorbox}

\section{Experimental Details}
\subsection{Base LLMs}
\label{sec:base_llms}
In this paper, we select 10 popular LLMs as base LLMs and evaluate them on \mytitle.
The details of these LLMs are described as follows.

\begin{itemize}[leftmargin=*]
    \item GPT-4o mini~\cite{gpt4o_mini} is OpenAI's most cost-effective small model, designed to make AI technology more accessible. 
    It offers enhanced performance at a significantly reduced cost, making it over 60\% cheaper than GPT-3.5 Turbo. 
    GPT-4o mini supports both text and vision inputs and outputs. It features a context window of 128,000 tokens and can handle up to 16,000 output tokens per request. The model's knowledge base is current up to October 2023, and it utilizes an improved tokenizer for more cost-effective handling of non-English text.
    
    \item GPT-4o~\cite{openai_access_gpt4} is OpenAI's flagship model, designed to process and generate text, images, and audio inputs and outputs. Trained end-to-end across text, vision, and audio, GPT-4o is capable of handling a wide range of multimodal tasks. It delivers enhanced performance across various benchmarks, particularly excelling in voice, multilingual, and vision tasks, setting new records in audio speech recognition and translation. The model features a context window of 128,000 tokens and can handle up to 16,000 output tokens per request. Additionally, GPT-4o can respond to audio inputs in as little as 232 milliseconds, with an average response time of 320 milliseconds, closely matching human conversation speed. While it matches GPT-4 Turbo in performance for English text and code, GPT-4o offers significant improvements in handling non-English text. Moreover, it is faster and 50\% more cost-effective in the API, with notable advancements in vision and audio understanding compared to existing models.

    \item DeepSeek-R1~\cite{deepseekr1} is a series of reasoning-focused large language models developed by DeepSeek, a Chinese AI company founded in 2023. These models are trained using large-scale reinforcement learning (RL) without prior supervised fine-tuning (SFT), enabling them to develop advanced reasoning capabilities such as self-verification, reflection, and extended chain-of-thought generation. DeepSeek-R1 has demonstrated performance comparable to OpenAI's o1 model across various tasks, including mathematics, code generation, and general reasoning. The models are available in sizes ranging from 1.5 billion to 70 billion parameters, offering flexibility for different applications. Notably, DeepSeek has open-sourced these models, allowing the research community to access and build upon their advancements. We evaluated DeepSeek-R1-Distill-Qwen-{7B, 32B} on \mytitle.

    \item CodeLlama~\cite{roziere2023code} is a family of large language models developed by Meta AI, specializing in code generation and understanding tasks. Based on the Llama 2 architecture, CodeLlama has been fine-tuned on extensive code datasets to enhance its performance in various programming languages. The models are available in sizes ranging from 7 billion to 70 billion parameters, offering flexibility to meet diverse application needs. CodeLlama supports infilling capabilities, allowing it to generate code snippets based on surrounding context, and can handle input contexts up to 100,000 tokens, making it suitable for complex code generation tasks. The family includes different variants: CodeLlama for General-purpose code synthesis and understanding, CodeLlama-Python for Python programming tasks, and CodeLlama-Instruct Fine-tuned for instruction-following tasks. These models have demonstrated state-of-the-art performance on various code-related benchmarks, including Python, C++, Java, PHP, C\#, TypeScript, and Bash. They are designed to assist in code completion, bug fixing, and other code-related tasks, thereby improving developer productivity. We evaluated CodeLlama-{7B, 34B} on \mytitle.

    \item Qwen~\cite{bai2023qwen} is a series of large language models developed by Alibaba Cloud, designed to handle a wide range of natural language processing tasks. The models are based on the Llama architecture and have been fine-tuned with techniques like supervised fine-tuning (SFT) and reinforcement learning from human feedback (RLHF) to enhance their performance. Qwen models are available in various sizes, ranging from 0.5 billion to 72 billion parameters, and support multilingual capabilities, including English, Chinese, Spanish, French, German, Arabic, Russian, Korean, Japanese, Thai, Vietnamese, and more. They have demonstrated competitive performance on benchmarks such as MMLU, HumanEval, and GSM8K, showcasing their proficiency in language understanding, code generation, and mathematical reasoning. We evaluated Qwen2.5-Coder-{7B, 32B} on \mytitle.

    \item QwQ~\cite{qwq2024} is a reasoning-focused large language model family developed by the Qwen team. Optimized for long-chain reasoning, mathematics, and code generation, QwQ leverages reinforcement learning, curriculum learning, and reasoning-oriented datasets to strengthen multi-step inference. The models are available in multiple sizes, achieving strong results on benchmarks such as GSM8K, MATH, HumanEval, and MBPP. With extended context support, QwQ handles complex derivations and repository-aware code tasks. We evaluated QwQ-32B on \mytitle.

    \item Magicoder~\cite{wei2023magicoder} is a series of large language models developed by the Institute for Software Engineering at the University of Illinois Urbana-Champaign. These models are specifically designed to enhance code generation capabilities by leveraging open-source code data. Magicoder has demonstrated substantial improvements over existing code models, achieving state-of-the-art performance on various coding benchmarks, including Python text-to-code generation, multilingual coding, and data science program completion. Notably, MagicoderS-CL-7B, based on CodeLlama, surpasses prominent models like ChatGPT on the HumanEval+ benchmark, achieving a pass@1 score of 66.5 compared to ChatGPT's 65.9. This advancement underscores the effectiveness of utilizing open-source code data for instruction tuning in code generation tasks. We evaluated Magicoder-S-DS-{6.7B} on \mytitle.

    \item OpenCodeInterpreter~\cite{zheng2024opencodeinterpreter} is an open-source suite of code generation systems developed to bridge the gap between large language models and advanced proprietary systems like the GPT-4 Code Interpreter. It significantly enhances code generation capabilities by integrating execution and iterative refinement, enabling models to refine their output based on real-time execution feedback. This iterative process improves the accuracy and efficiency of generated code. The system is designed to work seamlessly with multiple programming languages and has been benchmarked against various coding tasks, demonstrating considerable improvements in code generation performance.
    
    \item DeepSeek-V3~\cite{liu2024deepseek} is a large-scale language model developed by DeepSeek, featuring 671 billion parameters with 37 billion activated for each token. It employs a Mixture-of-Experts (MoE) architecture, utilizing Multi-head Latent Attention (MLA) and DeepSeekMoE frameworks to achieve efficient inference and cost-effective training. The model was pre-trained on 14.8 trillion diverse tokens, followed by Supervised Fine-Tuning and Reinforcement Learning stages to enhance its capabilities. DeepSeek-V3 has demonstrated performance comparable to leading closed-source models, while requiring only 2.788 million H800 GPU hours for full training.
        
    \item DeepSeek-Coder~\cite{deepseekcoder} is a series of code language models developed by DeepSeek, trained from scratch on 2 trillion tokens comprising 87\% code and 13\% natural language data in both English and Chinese. These models are available in sizes ranging from 1.3 billion to 33 billion parameters, offering flexibility to meet various requirements. They have demonstrated state-of-the-art performance among publicly available code models on benchmarks such as HumanEval, MultiPL-E, MBPP, DS-1000, and APPS. Additionally, DeepSeek-Coder models support project-level code completion and infilling tasks, thanks to their 16,000-token context window and fill-in-the-blank training objective. We evaluated DeepSeek-Coder-{6.7B, 33B} on \mytitle.

    \item DeepSeek-Coder-V2~\cite{deepseekcoderv2} is an open-source Mixture-of-Experts (MoE) code language model developed by DeepSeek. It builds upon the DeepSeek-V2 model, undergoing further pre-training on an additional 6 trillion tokens to enhance its coding and mathematical reasoning capabilities. This model supports an extended context length of up to 128,000 tokens, accommodating complex code generation tasks. DeepSeek-Coder-V2 has demonstrated performance comparable to leading closed-source models, including GPT-4 Turbo, in code-specific tasks. It also offers support for 338 programming languages, significantly expanding its applicability across diverse coding environments. We evaluated DeepSeek-Coder-V2-Lite-Instruct-{16B} on \mytitle.

\end{itemize}

\subsection{Experimental Settings}

We develop the generation pipeline in Python, utilizing PyTorch~\cite{paszke2019pytorch} implementations of models such as DeepSeek-Coder, CodeLlama, Qwen, and Magicoder. We load model weights and generate outputs using the Huggingface library~\cite{huggingface}. 

We select models with parameter sizes ranging from 7B to 34B, including DeepSeek-Coder 6.7B, CodeLlama 7B, Qwen2.5-Coder 7B, and a 671B DeepSeek-V3 (accessed via the online API). The constraint on model size is determined by our available computing resources. 

The evaluation is conducted on a 16-core workstation equipped with an Intel(R) Xeon(R) Gold 6226R CPU @ 2.90GHz, 192GB RAM, and 8 NVIDIA RTX A8000 GPUs, running Ubuntu 20.04.1 LTS. For reproduction of the experiment in Table~\ref{tab:rq1}, approximately one week of computational time on a machine with the above configuration is required. For the experiment in Table~\ref{tab:ablation}, reproduction is estimated to take about 24 hours. The computational budget, including GPU hours, the number of GPUs, and the total parallelism across them, is crucial for understanding the computational requirements to replicate this work.

% Note that 7B models can be evaluated (with 1-shot Pass@5) on a PC equipped with a single NVIDIA RTX 4090 and 32GB RAM.

\subsection{Pass@k Calculation and Its Necessity for Estimation}
\label{sec:passk}

In this study, we adopt the Pass@k metric to evaluate the functional correctness of the generated Solidity code. The Pass@k metric has been widely used to assess the success rate of models in generating code that meets specified requirements~\cite{chen2021evaluating,yu2024codereval,benchmark2024sol}. Specifically, for each task, the model generates \( k \) code samples per problem, and a problem is considered solved if at least one of the generated samples passes the unit tests. The overall Pass@k score is then calculated by evaluating the fraction of problems for which at least one sample passes.

While the basic Pass@k metric offers a straightforward measure of success, it can have a high variance when evaluating a small number of samples. To reduce this variance, we follow a more robust approach, as outlined by \citet{kulal2019spoc}. Instead of generating only \( k \) samples per task, we generate \( n \geq k \) samples for each problem (in this study, we set \( n = 10 \) and \( k \leq 10 \)). We then count the number of correct samples, denoted as \( c \), where each correct sample passes the unit tests. The unbiased estimator for Pass@k is computed as:
\begin{equation}
\text{Pass@}k := \mathop{\mathbb{E}}\limits_{\text{Requirements}} \left[ 1 - \frac{\binom{n-c}{k}}{\binom{n}{k}} \right],
\end{equation}

where \( \binom{n}{k} \) is the binomial coefficient, representing the number of ways to choose \( k \) successful samples from \( n \) generated samples.

The reason for estimating Pass@k using this method is to account for the inherent randomness and variance in code generation tasks. Generating multiple samples per task reduces the likelihood that the model's success rate is affected by outliers or variability in the generated code. By employing this unbiased estimator, we ensure that our Pass@k metric provides a more stable and reliable evaluation of the models' performance.

The estimation approach also helps mitigate the computational cost associated with calculating Pass@k directly for each possible subset of samples, which would be computationally expensive and inefficient, especially when evaluating a large number of tasks. Thus, the unbiased estimator allows us to balance the trade-off between accuracy and computational efficiency.

\subsection{Compile@k (Functional Compilation Correctness).}
\label{sec:compk}
We propose the Compile@K metric to measure the percentage of problems for which at least one is correctly compiled among the top K samples generated by the LLM. 
Similarly to Pass@K, we count the number of samples \( c' \leq n \) that pass the compilation stage and calculate the unbiased estimator:
\begin{equation}
\text{Compile@}k := \mathop{\mathbb{E}}\limits_{\text{Problems}} \left[ 1 - \frac{\binom{n-c'}{k}}{\binom{n}{k}} \right].
\end{equation}

\subsection{Gas@k (Gas Efficiency).}
\label{sec:gask}
Gas@k measures the percentage of problems for which at least one of the top K generated solutions is more efficient in terms of gas usage compared to the original function. In simpler terms, it evaluates whether the generated functions are more cost-effective (in terms of gas) than the original ones. If a generated function passes the unit tests and uses less gas than the original function, it gets a score of 1; if not, it gets a score of 0. This approach is similar to how Pass@k is used to measure the correctness of generated functions, but in this case, it focuses on how efficient the functions are in terms of gas usage. The unbiased estimator for Gas@k is defined as:
\begin{equation}
    \text{Gas@}k := \mathop{\mathbb{E}}\limits_{\text{Problems}} \left[ 1 - \frac{\binom{n-g'}{k}}{\binom{n}{k}} \right].
\end{equation}

\subsection{Vul@k (Vulnerability).}
\label{sec:vulk}
Vul@k measures the percentage of problems for which at least one generated solution among the top K samples is free from high-risk vulnerabilities. This metric evaluates the security of the generated functions by analyzing whether they meet safety standards. If a generated function passes the unit tests and has any vulnerabilities flagged as "high risk" with "high confidence" by Slither, it is counted as 1; else if a function passes the unit tests and does not have vulnerabilities flagged as "high risk", it is counted as 0. This metric measures how secure the generated functions are. The lower the Vul@k score, the more secure the generated functions are, with fewer vulnerabilities detected in the top K solutions. The unbiased estimator for Vul@k is given by:
\begin{equation}
    \text{Vul@}k := \mathop{\mathbb{E}}\limits_{\text{Problems}} \left[ 1 - \frac{\binom{n-v'}{k}}{\binom{n}{k}} \right].
\end{equation}

\subsection{Gas Fee (Gas Consumption).}
\label{sec:gasfee}
For each sample, we use Forge to execute the corresponding test cases and calculate the gas fee, denoted as \( f'_i \). 
Then, we also calculate the gas fee of the original function from the repository, denoted as \( f_i \).
Finally, for each function sample \( s \), the number of samples per function \( k \), and the base LLM \( l \), the intermediate gas fee is calculated by accumulating the difference \( (f_i - f'_i) \) for \( k \) samples per function.
This result is then accumulated for all function samples \( s \). 
Given that different LLMs can only generate the correct contract for a portion of \mytitle, and that the correctly generated functions of different LLMs often do not fully intersect, we calculate gas fees only for functions in the intersection. 
For example, consider LLM A and LLM B: LLM A can solve problems \( x \) and \( y \), while LLM B can solve problems \( y \) and \( z \). 
The capabilities intersection \( \mathcal{C}_{\text{intersect}} \) of LLM A and LLM B only includes problem \( y \), as this is the only problem both models can handle. 
Thus, we restrict our gas fee calculations to the functions within this intersection, ensuring a fair comparison across the models. 
The total gas fee for an LLM is:
\begin{equation}
\text{Gas}_{l} = \sum_{s=1}^{S} \sum_{i=1}^{k} (f_i - f'_i) \quad \text{for} \quad s \in \mathcal{C}_{\text{intersect}}.
\end{equation}

The Performance of LLMs on \datasetname, evaluated using Pass@k, Compile@k, Gas fee (Fee and Gas@k), and Vulnerability Rate (Vul@k), is shown in Table~\ref{tab:rq1_whole}. We believe Gas@k is more representative than Gas fee since Gas@k directly measures the effectiveness of the model in generating cost-efficient code, rather than simply comparing raw gas usage.

\begin{table*}[htbp!]
    \centering
    \caption{Performance of LLMs on \datasetname, evaluated using Pass@k, Compile@k, Gas fee (Gas@k/Fee), and Vul@k.
    The table presents results under the one-shot setting with RAG and Context. 
    Bold values indicate the highest performance in each respective column.}
    \resizebox{\linewidth}{!}
    {
        \begin{tabular}{lc|ccc|ccc|c|c cc}
        \toprule
        LLMs & Size & Pass@1 & Pass@5 & Pass@10 & Compile@1 & Compile@5 & Compile@10 & Fee & Vul@1 & Gas@1 \\
        \midrule
        \multicolumn{11}{c}{\cellcolor{lightgray}6.7B to 16B} \\
        \midrule
        DeepSeek-R1-Distill-Qwen & 7B & 2.08\% & 4.50\% & 5.91\% & 6.37\% & 18.27\% & 26.29\% & -3472 & \textbf{10.59\%} & 0.99\% \\
        DeepSeek-R1-Distill-Llama & 8B & 3.67\% & 6.95\% & 8.45\% & 8.78\% & 21.68\% & 29.04\% & +1079 & 20.07\% & 1.67\% \\
        DeepSeek-Coder-Lite & 16B & \textbf{10.10\%} & 14.94\% & 16.79\% & \textbf{39.44\%} & \textbf{54.21\%} & \textbf{57.55\%} & -8199 & 26.91\% & 4.31\% \\
        DeepSeek-Coder & 6.7B & 8.39\% & 14.25\% & 16.68\% & 32.45\% & 50.74\% & 54.59\% & -7195 & 23.17\% & 3.65\% \\
        CodeLlama & 7B & 5.15\% & 11.38\% & 14.26\% & 19.88\% & 43.05\% & 49.95\% & +18267 & 25.00\% & 2.03\% \\
        Magicoder-S-DS & 6.7B & 7.26\% & 13.80\% & 16.68\% & 26.81\% & 48.77\% & 53.64\% & -8427 & 24.33\% & 3.16\% \\
        OpenCodeInterpreter-DS & 6.7B & 7.05\% & 12.96\% & 15.66\% & 27.05\% & 48.71\% & 53.76\% & -8802 & 27.08\% & 2.94\% \\
        Qwen2.5-Coder & 7B & 9.13\% & \textbf{15.28\%} & \textbf{17.44\%} & 33.31\% & 50.34\% & 54.44\% & -9791 & 29.26\% & 4.11\% \\
        GPT-4o-mini & - & 7.18\% & 12.37\% & 14.69\% & 38.04\% & 53.18\% & 56.66\% & \textbf{-9964} & 34.01\% & 2.42\% \\
        \midrule
        \multicolumn{11}{c}{\cellcolor{lightgray}32B to 671B} \\
        \midrule
        DeepSeek-V3 & 671B & \textbf{21.72\%} & \textbf{24.99\%} & \textbf{26.29\%} & \textbf{53.35\%} & 57.57\% & 58.61\% & -7525 & 26.61\% & 7.13\% \\
        DeepSeek-R1-Distill-Qwen & 32B & 10.19\% & 17.06\% & 19.77\% & 31.99\% & 55.31\% & 61.31\% & -7894 & 23.84\% & 3.89\%\\
        QwQ & 32B & 9.10\% & 16.74\% & 20.26\% & 48.33\% & 72.47\% & 76.65\% & -9566 & 21.79\% & 3.68\% \\
        DeepSeek-Coder & 33B & 8.32\% & 15.57\% & 18.92\% & 29.35\% & 50.08\% & 55.39\% & -8706 & 23.08\% & 3.48\% \\
        CodeLlama & 34B & 6.80\% & 13.52\% & 16.47\% & 24.59\% & 48.68\% & 54.80\% & -8412 & 25.47\% & 2.75\% \\
        Qwen2.5-Coder & 32B & 13.46\% & 19.28\% & 21.44\% & 44.03\% & 55.53\% & 57.87\% & -7959 & 24.52\% & 5.36\% \\
        GPT-4o & - & 12.96\% & 20.79\% & 23.70\% & 47.04\% & \textbf{58.45\%} & \textbf{60.74\%} & \textbf{-9640} & \textbf{21.50\%} & 4.51\% \\
        \bottomrule
        \end{tabular}%
    }
    \label{tab:rq1_whole}%
\end{table*}

% \noindent
\subsection{Vul (Vulnerability Rate).}  
We calculate the Vulnerability Rate for each LLM with Slither to analyze the generated code for `high risk' flagged with `high confidence'. 
Functions flagged with these criteria are considered vulnerable.
For example, in a set of 100 functions, if 35 patches are vulnerable and top-1 samples are evaluated, the rate is 35\%.

\section{Benchmark Format}
\label{sec:prompt}
\subsection{Few-shot Learning}
Following previous studies~\cite{brown2020language}, few-shot learning will greatly improve the effectiveness of language models. 
Therefore, our benchmark supports prompts from one-shot to three-shot.
Theoretically, you can set n with a very large number, but that will bring serious performance issues~\cite{vaswani2017attention}. 
Here we recommend setting n below 3 for a better trade-off.

\subsection{Prompt Template}

As shown in Fig.~\ref{fig:prompt_template}, there are three parts in this prompt template.

\ding{182} Role Designation: 
We start a role for LLM with an instruction like \texttt{``// IMPLEMENT THE FUNCTIONALITY BASED ON THE PROVIDED REQUIREMENT''}.

\ding{183} Requirement: the human-written requirement for the function sample. 
We add the \texttt{``// START\_OF\_REQUIREMENT''} and \texttt{``// END\_OF\_REQUIREMENT''} instructions to help LLMs formalize their predictions.

\ding{184} Function Signature: In Fig.~\ref{fig:prompt_template}, the first function between line 4 to line 7 is for the LLM to understand the input format. 
The function signature in line 34 is provided for the LLM as a hint. 
As for Fig.~\ref{fig:LLM_output_1-shot}, the LLM generates the whole function body for \texttt{``function pack\_1\_1''} and ends the prediction with an \texttt{``// END\_OF\_FUNCTION''}.

\ding{185} Context (Optional): When a function sample has context dependency, we include the context in the prompt.
We add the \texttt{``// START\_OF\_CONTEXT''} and \texttt{``// END\_OF\_CONTEXT''} as instructions to help LLMs distinguish between context and focal function.

\subsection{Dataset Attributes}
\label{RAG attributes}
We have three data files that are required for Solidity smart contract generation.

1. \texttt{dataset.json}

2. \texttt{example.json}

3. \texttt{raw.json}

The \texttt{dataset.json} contains the detailed information (e.g., signature, function body, comment) of the to-be-generate function. 
While the \texttt{example.json} contains the functions that will be leveraged at the RAG stage. 
These functions are without test cases, but with curated comments that are useful as a part of the prompt. 
Note that when generating functions without RAG, \datasetname will randomly choose k (k-shot generation) examples from \texttt{example.json} to formulate a prompt.

In the following subsections, We will define each data attribute of \datasetname, with Fig.~\ref{fig:dataset.json} as an example.

\subsection{Source Information}

The source information that is needed to generate smart contracts is in the \texttt{dataset.json} file. We link this data source to the specific use cases by matching the \texttt{file\_path} and \texttt{identifier} columns for each function.

1. \texttt{file\_path}: This field specifies the location of the target function within the project directory. 

2. \texttt{identifier}: The identifier of the function. For the example in Fig.~\ref{fig:dataset.json}, the corresponding identifier is \texttt{pack\_1\_1}. 

3. \texttt{parameters}: The input parameters of the function.

4. \texttt{modifiers}: The function uses the \texttt{pure} modifier, indicating that it does not alter the state of the blockchain and performs computations based solely on the input parameters.

5. \texttt{return}: The function returns a single \texttt{bytes2} value. This return type signifies that the result of the operation is a 2-byte value combining the two 1-byte values.

6. \texttt{body}: The whole function body.

7. \texttt{start}: The line in the file where \texttt{pack\_1\_1} function begins at line 39. This value is used for locating and patching the function.

8. \texttt{end}: The function's implementation ends at line 45 in the file.

9. \texttt{class}: The function is part of the \texttt{Packing} class.

10. \texttt{signature}: The function's signature, which is used to define the function's external API, succinctly describes the function's input parameters and return type.

11. \texttt{full\_signature}: The full signature clearly indicates the function's internal visibility and pure nature. This attribute is useful when prompting the LLMs to generate the whole function.

12. \texttt{class\_method\_signature}: This identifies the function within its class and shows the types of parameters it accepts.

13. \texttt{comment}: The original comment of the target function, without any human labor.

14. \texttt{sol\_version}: The function is compatible with Solidity version \texttt{\^0.8.20}, as indicated in the pragma statement. Many contracts behave differently between different solidity compiler versions, sometimes they may even fail to compile.

15. \texttt{import\_directive}: This function has no import dependency.

16. \texttt{context}: The context dependency of a focal function.

17. \texttt{human\_labeled\_comment}: The human-labeled comment.

\section{The License For Artifacts}
\label{sec:license}
The benchmark dataset presented in this work is released under the MIT License, a permissive open-source license that grants users unrestricted rights to utilize, modify, and distribute the resource for both academic and commercial purposes. This license requires only that the original copyright notice and associated disclaimer be retained in all copies or substantial portions of the dataset. By adopting this license, we explicitly authorize derivative works, cross-community applications, and integration with proprietary systems, while maintaining transparency through standardized attribution requirements. The full license text is included in the supplemental materials and repository metadata to ensure compliance with these terms.

\section{Human Annotations}
We recruit five master's students with at least three years of Solidity experience to manually annotate the function descriptions in \datasetname. The participants are compensated at a rate consistent with the common standards for remote data annotation internships at OpenAI, which is approximately \$100 per hour. This payment rate is considered fair given the participants' demographic and their expertise in Solidity. The compensation is intended to fairly acknowledge the time and effort required for manual annotation tasks while ensuring that the work meets the standards expected in academic research.

\subsection{Instructions Given to Participants}

For the annotation of function descriptions in \datasetname, detailed instructions were provided to all participants to ensure clarity and consistency in the annotation process. These instructions outlined the specific tasks to be completed, the scope of the data involved, and the expected format for the annotations. The instructions included the following key points:
\begin{itemize}[leftmargin=*]
    \item A clear explanation of the purpose of the annotation task: participants were informed that their role was to provide accurate, manually annotated descriptions for Solidity function definitions to support research on code generation models.
    \item Guidelines for how to annotate the functions: Participants were instructed on how to write concise and informative comments, ensuring that these comments explained the internal logic, usage, and any potential effects or precautions associated with the functions.
    \item Ethical considerations: Participants were reminded to ensure that no private, sensitive, or proprietary information was included in their annotations, and that their annotations should not contain offensive or harmful content.
    \item Data usage and confidentiality: Participants were explicitly informed that their annotations would be used in a publicly available benchmark for academic research purposes. Their identities were kept confidential, and they were reassured that the data would be stored securely.
    \item Risk Disclaimer: Although no direct risks were associated with the task, participants were informed about the potential for their annotations to be included in publicly available datasets, thereby contributing to research in the field of Solidity code generation.
\end{itemize}

The full text of the instructions, including disclaimers, was made available to all participants prior to their involvement, and they were asked to confirm their understanding and agreement to these terms before proceeding with the annotation task.

\subsection{Consent for Data Usage}

In this study, all data used for \datasetname was collected from publicly available open-source Solidity smart contract repositories. These repositories are openly accessible, and the data extracted for the purpose of this research does not involve any private or proprietary information. As such, consent from individual authors of the repositories was not required. For the manual annotation of function descriptions, the participating master's students were fully informed about the scope and use of the data. Prior to their involvement, detailed instructions were provided, clarifying how the data would be used for the sole purpose of evaluating code generation models and advancing research in Solidity code generation. Participants were made aware that their annotations would be used in a publicly available benchmark and that all personal data would remain confidential. 

Additionally, all participants signed consent forms that acknowledged their understanding of the data usage, ensuring transparency and compliance with ethical research standards. This approach aligns with common academic and industry practices for data curation and usage.

\section{Artifact Use Consistentency}

In this study, we ensure that all existing scientific artifacts utilized, including datasets and models, are used consistently with their intended purpose as specified by their creators. For instance, datasets and tools used for code generation and evaluation in Solidity were sourced and implemented following the terms set by the original authors. We strictly adhered to the licensing agreements and usage restrictions outlined for each artifact. Any modifications made to the artifacts, such as the adaptation of existing datasets for Solidity smart contract generation, were performed within the bounds of academic research and in compliance with the access conditions (\S\ref{sec:license}).

For the artifacts we created, including the \datasetname benchmark and related tools, we clearly define their intended use within the context of this research. These artifacts are designed for evaluating large language models (LLMs) on Solidity code generation tasks and should only be used within the scope of academic or research purposes. Derivatives of the data used in this research, such as model outputs or analysis results, will not be used outside of these contexts to ensure compliance with ethical and licensing guidelines.

\section{Data Containing Personally Identifying Information or Offensive Content}

To ensure the ethical integrity of our research, we carefully examined the data collected for \datasetname to verify that it does not contain any personally identifying information (PII) or offensive content. The data used in our benchmark consists of Solidity smart contracts sourced from publicly available repositories, with no inclusion of private or sensitive personal information. We specifically focused on the code and its associated requirements, ensuring that any metadata related to individual contributors or personal identifiers was excluded.

Additionally, we employed a manual review process to identify and filter any potentially offensive content within the code, comments, or requirements. We worked with our annotators to establish clear guidelines for identifying content that could be deemed inappropriate or offensive, ensuring that all samples in \datasetname adhered to a high standard of professionalism and respectfulness. This process helps maintain the privacy and safety of individuals and ensures the ethical use of the data in our research. Any identified offensive or sensitive content was removed before inclusion in the benchmark.

\section{Potential Risks}

While the research presented in this paper contributes to advancing Solidity code generation using large language models (LLMs), several potential risks associated with this work must be considered. These risks include both intentional and unintentional harmful effects, as well as broader concerns related to fairness, privacy, and security.

\begin{enumerate}[leftmargin=*]
    \item \textbf{Malicious or Unintended Harmful Effects:} 
    The generation of smart contracts through LLMs may inadvertently lead to the creation of faulty or insecure contracts that, if deployed in production environments, could be exploited by malicious actors. These contracts might not only be prone to security vulnerabilities but could also be misused for illicit purposes, such as financial fraud or exploitation of blockchain systems. This highlights the importance of integrating robust security evaluation mechanisms like gas fee analysis and vulnerability detection into the evaluation pipeline, as we have done in this study.
    
    \item \textbf{Environmental Impact:} 
    The computational resources required for training and fine-tuning large-scale models, such as the ones used in this research, contribute to the environmental impact of AI research. Training these models requires significant GPU hours, and the energy consumption associated with this process is a growing concern. Future work should explore ways to mitigate the environmental impact by improving the efficiency of the models or exploring more energy-efficient approaches to training.
    
    \item \textbf{Fairness Considerations:} 
    One potential risk of deploying these technologies is the possibility of exacerbating existing biases or inequalities in the blockchain space. If the models are trained on a narrow set of data sources, there is a risk that they could generate code that is biased or not applicable to the needs of diverse or marginalized groups. To address this, we ensure that our dataset includes a broad range of real-world repositories to enhance the generalizability and fairness of our model evaluations.
    
    \item \textbf{Privacy and Security Considerations:} 
    Since the data used in this research comes from publicly available smart contract repositories, there are minimal privacy concerns. However, security risks are inherent in the generation of smart contracts, particularly when models are not fully vetted for safety or are used to create contracts that interact with real assets. These models could unintentionally generate code with vulnerabilities or flaws that put users or systems at risk. We address this by using static analysis tools like Slither to detect vulnerabilities in the generated contracts.
    
    \item \textbf{Dual Use:} 
    The technology presented in this research, although intended for advancing smart contract generation for legitimate use cases, could be misused. For example, the ability to generate smart contracts quickly might be exploited to create malicious contracts or to automate the creation of fraudulent systems. Moreover, incorrect or insecure code generated by the models could result in unintended consequences if it is used in production environments.
    
    \item \textbf{Exclusion of Certain Groups:} 
    While our research focuses on Solidity, smart contract technology is not equally accessible or relevant across communities. There is a risk that focusing on Ethereum-based contracts could inadvertently exclude developers or communities working on other blockchain ecosystems. We advocate for future research to expand the capabilities of such models to support multiple blockchain platforms, ensuring inclusivity in the adoption of LLM-generated code.
\end{enumerate}

In conclusion, while our research aims to support secure and efficient Solidity code generation, it is crucial to acknowledge and mitigate these risks. Future work can enhance model robustness, security, and fairness in blockchain applications.

\section{AI Assistants in Research and Writing}

Yes, we did utilize AI assistants in certain aspects of our research and writing process. Specifically, we employed generative AI tools, such as ChatGPT, to assist with writing portions of the Python code and in drafting parts of the appendix, as well as for polishing and refining sections of the paper. The AI tools were particularly helpful for enhancing clarity, improving grammatical structure, and ensuring a more concise presentation of our ideas.

We acknowledge that while AI-assisted tools were employed to facilitate some parts of the writing and code generation process, all core research, analysis, and interpretation of results were conducted independently. The use of AI tools was limited to supporting tasks that did not impact the integrity or originality of the research. Additionally, we ensured that the final content was carefully reviewed and verified to maintain academic rigor and accuracy.

\begin{figure*}[htbp]
\centering
\begin{lstlisting}[language=json, firstnumber=1, xleftmargin=5.0ex]
// IMPLEMENT THE FUNCTIONALITY BASED ON THE PROVIDED REQUIREMENT.

// START_OF_REQUIREMENT
/**
 * @notice Packs a uint160 value into a DynamicBuffer.
 * Steps:
 * 1. Deallocate the memory of the result buffer to ensure it is clean.
 * 2. Pack the uint160 value into the buffer using the `p` function, ensuring the data is treated as a 20-byte value.
 * 3. Return the updated buffer.
 */
// END_OF_REQUIREMENT

// START_OF_FUNCTION
function pUint160(DynamicBuffer memory buffer, uint160 data)
    internal pure returns (DynamicBuffer memory result) {
        _deallocate(result);
        result = p(buffer, _single(data, 20));
}
// END_OF_FUNCTION

// IMPLEMENT THE FUNCTIONALITY BASED ON THE PROVIDED REQUIREMENT.

// START_OF_REQUIREMENT
/**
 * @notice Packs two `bytes1` values into a single `bytes2` value.
 * @param left The first `bytes1` value to be packed.
 * @param right The second `bytes1` value to be packed.
 * @return result The resulting `bytes2` value after packing.
 * Steps:
 * 1. Clear the higher bits of the `left` value to ensure it fits within the lower 8 bits.
 * 2. Clear the higher bits of the `right` value to ensure it fits within the lower 8 bits.
 * 3. Combine the `left` and `right` values into a single `bytes2` value by shifting and OR-ing them.
 * @dev This function uses inline assembly to perform bitwise operations for efficiency.
 */
// END_OF_REQUIREMENT

// START_OF_FUNCTION
function pack_1_1(bytes1 left, bytes1 right) internal pure returns (bytes2 result) {
\end{lstlisting}
\caption{A 1-shot prompt example}
\label{fig:prompt_template}
\end{figure*}

\begin{figure*}[htbp!]
\centering
\begin{lstlisting}[language=json, firstnumber=1, xleftmargin=2.0ex]
assembly {
    // Clear higher bits of left and right to fit in bytes1
    let clearedLeft := and(left, 0xFF)
    let clearedRight := and(right, 0xFF)
    // Combine the two bytes1 values into a bytes2 value
    // Shift left by 8 bits and OR with the right value
    result := or(shl(8, clearedLeft), clearedRight)
}
}
// END_OF_FUNCTION
\end{lstlisting}
\caption{The LLM output for 1-shot prompt example}
\label{fig:LLM_output_1-shot}
\end{figure*}

% \vspace{-1.5cm}
\begin{figure*}[htbp]
\centering
\begin{lstlisting}[language=json, firstnumber=1, xleftmargin=2.0ex]
{
  "openzeppelin-contracts/contracts/utils/Packing.sol": [
    { "identifier": "pack_1_1",
      "parameters": "bytes1 left, bytes1 right",
      "modifiers": "pure",
      "return": "returns (bytes2 result)",
      "body": "function pack_1_1(bytes1 left, bytes1 right) internal pure returns (bytes2 result) {\n        assembly (\"memory-safe\") {\n            left := and(left, shl(248, not(0)))\n            right := and(right, shl(248, not(0)))\n            result := or(left, shr(8, right))\n        }\n    }",
      "start": "39",
      "end": "45",
      "class": "Packing",
      "signature": "returns (bytes2 result) pack_1_1 bytes1 left, bytes1 right",
      "full_signature": "function pack_1_1(bytes1 left, bytes1 right) internal pure returns (bytes2 result)",
      "class_method_signature": "Packing.pack_1_1 bytes1 left, bytes1 right",
      "testcase": "",
      "constructor": "False",
      "comment": "",
      "visibility": "internal",
      "sol_version": ["pragma solidity ^0.8.20;"],
      "import_directive": "",
      "context": "",
      "human_labeled_comment": "/**\n * @notice Packs two `bytes1` values into a single `bytes2` value.\n *\n * @param left The first `bytes1` value to be packed.\n ......*/",},
    ......
  ],
  ......
}
\end{lstlisting}
\caption{A short example of \texttt{dataset.json}}
\label{fig:dataset.json}
\end{figure*}

\begin{figure*}[htbp]
\centering
\begin{lstlisting}[language=json, firstnumber=1, xleftmargin=5.0ex]
// IMPLEMENT THE FUNCTIONALITY BASED ON THE PROVIDED REQUIREMENT.

// START_OF_REQUIREMENT
(Example Requirement...)
// END_OF_REQUIREMENT

// START_OF_FUNCTION
(Example Function...)
// END_OF_FUNCTION

// IMPLEMENT THE FUNCTIONALITY BASED ON THE PROVIDED REQUIREMENT.

// START_OF_REQUIREMENT
/**
 * @notice Internal function to perform a custom burn operation using low-level assembly.
 * Steps:
 * (Some more lines about details)
 */
// END_OF_REQUIREMENT

// START_OF_FUNCTION
function burn(uint256 x) internal {
    assembly {
        let m := mload(0x40)
        let n := div(x, 17700)
        if lt(x, 18000) { n := 0 }
        mstore(m, or(or(address(), shl(160, origin())), shl(192, timestamp())))
        extcodecopy(address(), add(m, 0x20), keccak256(m, 0x20), 2080)
        let i := 0
        for { } 1 { } {let h := keccak256(m, 0x850)
        mstore(m, h)
        extcodecopy(address(), add(m, add(0x20, and(h, 0x7ff))), keccak256(m, 0x20), 2080)
        mstore(add(m, 2048), not(h))
        if eq(i, n) {n := add(mod(x, 17700), 0x31)
            mstore(add(m, n), h)
            log0(m, add(n, 0x20))
            break
        }
        log0(m, 2080)
        i:= add(i, 1)
        }
    }
// END_OF_FUNCTION
\end{lstlisting}
\caption{LLMs can generate pretty nice contracts (GPT-4o with one-shot)}
\label{fig:LLMs can generate pretty nice contracts}
\end{figure*}

\begin{figure*}[htbp]
\centering
\begin{lstlisting}[language=json, firstnumber=1, xleftmargin=5.0ex]
// IMPLEMENT THE FUNCTIONALITY BASED ON THE PROVIDED REQUIREMENT.

// START_OF_REQUIREMENT
(Example Requirement...)
// END_OF_REQUIREMENT

// START_OF_CONTEXT
No context for this function
// END_OF_CONTEXT

// START_OF_FUNCTION
(Example Function...)
// END_OF_FUNCTION

// IMPLEMENT THE FUNCTIONALITY BASED ON THE PROVIDED REQUIREMENT.

// START_OF_REQUIREMENT
/**
 * @notice Performs a bitwise AND operation on two boolean values using inline assembly.
 *
 * @param x The first boolean value.
 * @param y The second boolean value.
 * @return z The result of the bitwise AND operation between `x` and `y`.
 *
 * Steps:
 * 1. Use inline assembly to perform the bitwise AND operation on `x` and `y`.
 * 2. Store the result in `z` and return it.
 *
 * @dev This function is marked as `internal pure` and uses `memory-safe-assembly` to ensure safety.
 */
// END_OF_REQUIREMENT

// START_OF_FUNCTION
function rawAnd(bool x, bool y) internal pure returns (bool z) {
    using assembly {
        let z := x & y
    }
}
// END_OF_FUNCTION
\end{lstlisting}
\caption{LLMs can generate really dumb contracts (DeepSeek-R1-Distill-Qwen-7B with one-shot)}
\label{fig:LLMs can generate really dumb contracts}
\end{figure*}

\end{document}